\documentclass{ws-ijmpcs}
\usepackage{verbatim}

\begin{document}

\markboth{J. C. Algaba}
{Probing Helical B-Fields in AGN by RM Gradients}

%
\catchline{}{}{}{}{}
%

\title{PROBING HELICAL MAGNETIC FIELDS IN AGN BY ROTATION MEASURE GRADIENTS STUDIES}

\author{J. C. ALGABA}

\address{Academia Sinica, Institute of Astronomy and Astrophysics\\
P.O. Box 23-141, Taipei 10617, 
Taiwan, R.O.C.\\
algaba@asiaa.sinica.edu.tw}



\maketitle

\begin{history}
\received{Day Month Year}
\revised{Day Month Year}
\end{history}

\begin{abstract}
One of the tools that can provide evidence about the existence of helical magnetic fields in AGN is the observation of rotation measure gradients across the jet. Such observations have been previously made successfully, proving that such gradients are far from being rare, but common and typically persistent over several years, although some of them may show a reversal in the direction along the jet. Further studies of rotation measure gradients can help us in our understanding of the magnetic field properties and structure in the base of the jets. We studied Very Long Baseline Array (VLBA) polarimetric observations of 8 sources consistent of some quasars and BL Lacs at 12, 15, 22, 24 and 43 GHz and we find that all but two sources show indications of rotation measure gradients, either parallel or perpendicular to the jet. We interpret gradients perpendicular to the jet as indications of the change of the line of sight of the magnetic field due to its helicity, and gradients parallel to the jet as the decrease of magnetic field strength and/or electron density as we move along the jet. When comparing our results with the literature, we find temptative evidence of a rotation measure gradient flip, which can be explained as a change of the pitch angle or jet bending.

\keywords{Keyword1; keyword2; keyword3.}
\end{abstract}

\ccode{PACS numbers: 11.25.Hf, 123.1K}

\section{Rotation Measure and Helical Magnetic Fields}
There is evidence that helical magnetic fields seems to be present in, at least, a large fraction of AGN. Gabuzda et.\ al (2006)\cite{Gabuzda2006} summarizes different observational polarization tests such as spine-sheath structures or the increase of polarization in the edges of the jet in order to distinguish between helical magnetic fields and alternative explanations. Although the mentioned effects could be caused by the combination of shocks and interactions with the media, helical magnetic fields alone can explain all of them. Even more, there is a smoking gun for testing the presence of helical magnetic fields: rotation measures. As radiation passes through a magnetized plasma, the polarization vectors rotate an angle proportional to the wavelength squared and the rotation measure, given by
\begin{equation}
RM\propto\int n_e B_{LOS}dl 
\end{equation}
The line of sight is very important: as we move along a helical magnetic line across the jet, the line of sight changes towards us, perpendicular to us, and finally away from us. In the simplest case, we should be able to identify RM gradients across the jet, due only to the change of the direction of the B field along the line of sight. This model gets more complicated as we consider viewing angles or magnetic/particle density gradients, but the key point is that a gradient will still exist and, if it spans from positive to negative values, it can only be due to a helical B field, as particle gradients or interaction with the media cannot reproduce these gradients\cite{Asada2002}.

\section{Observations and Results}
We obtained VLBA polarimetric observations at 12, 15, 22, 24 and 43 GHz of 8 sources, including 6 quasars and 2 BL Lac objects. The data reduction and imaging were done using standard techniques and rotation measure maps were derived for all the sources. We found clear indications of RM gradients in 6 of our sources (5 quasars and 1 BL Lac object). We present here two examples of the gradients found in our sample. A more exhaustive analysis will be given elsewhere\cite{Algaba2012}.

\subsection{0906+430}
\begin{figure}[pb]
\vspace{-0.5cm}
\centerline{\psfig{file=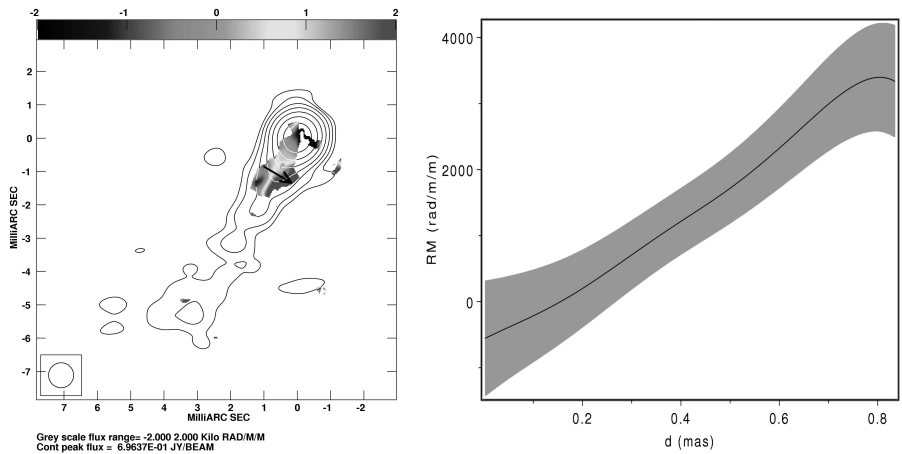,width=12cm}}
\caption{Left: 0906+430 rotation measure map. Right: slice of RM taken along the black solid arrow in the map. \label{f1}}
\end{figure}

In this source (see Fig. \ref{f1}) we find a clear RM gradient across the jet spanning along 2 mas ($\sim14$ pc), in agreement with Ventury \& Taylor (1999)\cite{VenturiTaylor1999}. Our RM values are higher, due to the higher frequencies used, in agreement with $RM\propto\nu^a$ (see Jorstad et.\ al 2007\cite{Jorstad2007}), with $a\sim3$ for this source. Both observations are in agreement with the assumption of a long-lived helical magnetic field driving the jet.

\subsection{0256+075}
\begin{figure}[pb]
\vspace{-0.5cm}
\centerline{\psfig{file=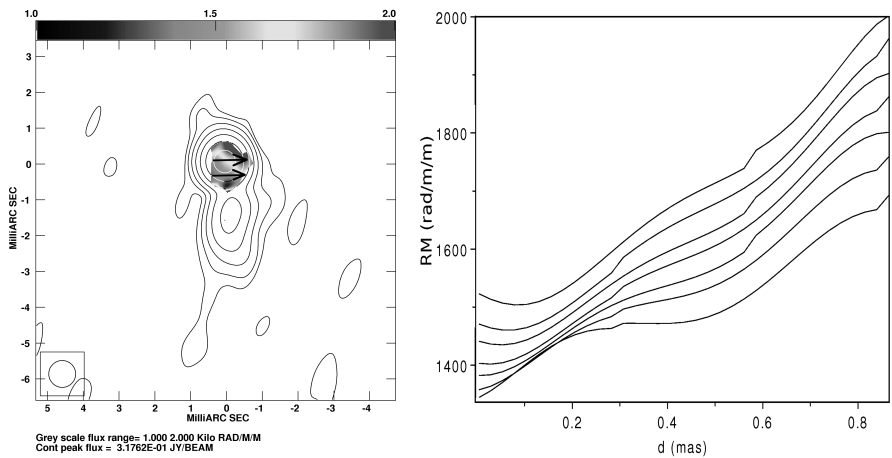,width=12cm}}
\caption{Left: 0256+075 rotation measure map. Right: series of slices of RM taken between the two black solid arrows in the map. \label{f2}}
\end{figure}

The RM gradient we find in this source (see Fig. \ref{f2}) is forming an angle of roughly $45^{\circ}$ with the jet. In Fig. \ref{f2}, right we show the evolution of the RM gradient as we move along the jet: we have obtained several RM slices across the jet between the two thick horizontal lines indicated in the RM map and we have plotted them together. Although the RM range covered is roughly of the same magnitude of the estimated errors, this provides us with information about i) the stability of the gradient along the jet and ii) the decrease of the absolute value of the RM as we move further from the inner region.

The behaviour of this RM gradient can be explained by the combination of these two factors: 1) the transverse RM due to the change of the line of sight of the magnetic field and 2) decrease of magnetic and/or particle density as we move out from the inner regions.

If we compare our results with these from Mahmud et.\ al (2009)\cite{Mahmud2009}, our beamsize is smaller as we are probing at higher frequencies. Hence, the resolved structure we observe corresponds to Mahmud's VLBA core. Again, we find higher RM values, with $a\sim2$ for this source, but we do not observe any RM indications of a gradient flip that was observed by them.

Mahmud et. al considered the magnetic tower model as an explanation for the RM gradient flip they observed. In this model, the magnetic lines anchored to the inner regions would come back and, as they are torqued by the motion of the disk, a double helix, one nested within the other and with opposite sign, would result. Hence, depending on which one dominates, we would have a gradient in one direction or the other. However, we consider this explanation to be unlikely, as no reversal of the dominant component is expected\cite{Broderick2010}, although it cannot be completely ruled out.

Our gradient closer to the base of the jet seems to have an opposite direction than the innermost one from Mahmud et. al. This could be explained in two different ways: 1) The gradients flip several times along the jet due to torsions, change of the helicity or jet bending\cite{O'Sullivan2006} or 2) time evolution of RM features (and hence, time evolution of physical properties of the jet and/or Faraday screen). More observations of this source with multiple resolutions are necessary in order to discriminate among these possibilities.

\section{Conclusions}

We present two examples of observed RM gradients in our sample constituted by blazars and QSO. We interpret them as the combination of the transverse gradient due to the action of the changing line of sight of a helical magnetic fields plus the particle/magnetic density decay as we move away from the inner region.

When we compare our RM gradient in 0256+075 with previous results, we find out a possible multiple flip of the RM gradient. To explain this we propose either a time evolution of the gradient or else changes of the line of sight of the magnetic field, due to pitch angle variations or jet bending, for example. We consider magnetic tower models unlikely.

We find an increase of the RM with frequency when we compare with previous results, consistent with $RM\propto\nu^a$, with $a\sim2$ for 0256+075 and $a\sim3$ for 0906+430.


\begin{thebibliography}{0}    

\bibitem{Gabuzda2006}D. C. Gabuzda, in {\it Proceedings of the 8th European VLBI Network Symposium on The role of VLBI in the Golden Age for Radio Astronomy and EVN Users Meeting}. September 26-29, 2006. Torun, Poland. PoS(8thEVN)011
\bibitem{Asada2002}K. Asada, M. Inoue, Y. Uchida, S. Kameno, K. Fujisawa, S.  Iguchi, M. Mutoh, {\it Pub. Astron. Soc. Japan} {\bf 54}, 39 (2002).
\bibitem{Algaba2012}J. C. Algaba, High Frequency VLBI Rotation Measures of 8 AGN, to appear in {\it Month. Not. Royal Astron. Soc.}
\bibitem{VenturiTaylor1999}T. Venturi, G. B. Taylor, {\it Astron. Journal} {\bf 118}, 1931 (1999).
\bibitem{Jorstad2007}S. G. Jorstad et al., {\it Astron. Journal} {\bf 134}, 799 (2007)
\bibitem{Mahmud2009}M. Mahmud, D. C. Gabuzda, in {\it Proceedings of the 9th European VLBI Network Symposium on The role of VLBI in the Golden Age for Radio Astronomy and EVN Users Meeting}. September 23-26, 2008. Bologna, Italy. PoS(IX EVN Symposium)011
\bibitem{Broderick2010}A. E. Broderick, J. C. McKinney, {\it Astron. Journal} {\bf 725}, 750 (2010)
\bibitem{O'Sullivan2006}S. O. P. O'Sullivan, D. C. Gabuzda, in {\it Proceedings of the 8th European VLBI Network Symposium on The role of VLBI in the Golden Age for Radio Astronomy and EVN Users Meeting}. September 26-29, 2006. Torun, Poland. PoS(8thEVN)014

\end{thebibliography}
\end{document}